\documentclass[onecolumn,showpacs,12pt]{revtex4}
\usepackage{graphicx}
\usepackage{dcolumn}
\usepackage{bm}
\begin{document}
\newcommand{\hs}{\hspace*{0.5cm}}
\newcommand{\vs}{\vspace*{0.5cm}}
\newcommand{\be}{\begin{equation}}
\newcommand{\ee}{\end{equation}}
\newcommand{\bea}{\begin{eqnarray}}
\newcommand{\eea}{\end{eqnarray}}
\newcommand{\ben}{\begin{enumerate}}
\newcommand{\een}{\end{enumerate}}
\newcommand{\bde}{\begin{widetext}}
\newcommand{\ede}{\end{widetext}}
\newcommand{\nn}{\nonumber}
\newcommand{\crn}{\nonumber \\}
\newcommand{\Tr}{\mathrm{Tr}}
\newcommand{\non}{\nonumber}
\newcommand{\noi}{\noindent}
\newcommand{\al}{\alpha}
\newcommand{\la}{\lambda}
\newcommand{\bet}{\beta}
\newcommand{\ga}{\gamma}
\newcommand{\va}{\varphi}
\newcommand{\om}{\omega}
\newcommand{\pa}{\partial}
\newcommand{\+}{\dagger}
\newcommand{\fr}{\frac}
\newcommand{\bc}{\begin{center}}
\newcommand{\ec}{\end{center}}
\newcommand{\Ga}{\Gamma}
\newcommand{\de}{\delta}
\newcommand{\De}{\Delta}
\newcommand{\ep}{\epsilon}
\newcommand{\varep}{\varepsilon}
\newcommand{\ka}{\kappa}
\newcommand{\La}{\Lambda}
\newcommand{\si}{\sigma}
\newcommand{\Si}{\Sigma}
\newcommand{\ta}{\tau}
\newcommand{\up}{\upsilon}
\newcommand{\Up}{\Upsilon}
\newcommand{\ze}{\zeta}
\newcommand{\ps}{\psi}
\newcommand{\Ps}{\Psi}
\newcommand{\ph}{\phi}
\newcommand{\vph}{\varphi}
\newcommand{\Ph}{\Phi}
\newcommand{\Om}{\Omega}

\title{ \bf
Electroweak phase transition in the reduced minimal 3-3-1 model}
\author{Vo Quoc  Phong\footnote{vqphong@hcmus.edu.vn}}
\affiliation{  Department of Theoretical Physics, Ho Chi Minh City
University of Science, Vietnam}
\author{Hoang Ngoc Long\footnote{hnlong@iop.vast.ac.vn}}
\affiliation{Institute of Physics, Vietnamese  Academy of Science and Technology,
10 Dao Tan, Ba Dinh, Hanoi, Vietnam}
\author{Vo Thanh Van\footnote{vtvan@hcmus.edu.vn}}
\affiliation{  Department of Theoretical Physics, Ho Chi Minh City
University of Science, Vietnam}

\date{\today}

\begin{abstract}
The electroweak phase transition is considered  in  framework of the 
reduced minimal 3-3-1 model (RM331). 
Structure of phase transition in this model is divided into two periods. The first period 
 is the phase transition $SU(3) \rightarrow  SU(2)$
 at TeV scale and the second one
  is $SU(2) \rightarrow U(1)$, which is  the like-Standard Model
   electroweak phase transition. When mass of the neutral Higgs boson ($h_1$) 
   is taken to be equal to the LHC value:  $m_{h_1}=125$ GeV, then 
   these phase transitions are the first order phase transitions,  
   the mass of $Z_2$ is about $4.8$ TeV;
   and we find  the region of parameter space with the first order 
   phase transition at $v_{\rho_0}=246$ GeV scale,
    leading to an effective   potential, where mass of the  charged  
    Higgs boson is in range of  $3.258 \, \mathrm{TeV}  < m_{h_{++}} <   19.549 \, \mathrm{TeV}$. 
    Therefore, with this approach, new bosons are the triggers of the first
   order electroweak phase transition with significant implications for 
   the viability of electroweak baryogenesis scenarios.
\end{abstract}


\maketitle


\section{Introduction}
\label{secInt}

Electroweak phase transition (EWPT) is a type of symmetry-breaking phase transition 
which plays an important role at early stage of expanding universe. Particularly, the EWPT is important to
 explaining baryon asymmetry of our Universe. As proposed by Sakharov \cite{mkn}, 
 three necessary conditions that a baryon-generating interaction in a theoretical model must satisfy to 
 produce an excess of baryons over antibaryons are: baryon number violation, C and CP violations, 
 and deviation from thermal equilibrium \cite{mkn}.

If baryon number (B) is conserved and is equal to zero, it will equal to zero forever. 
In contrast, B will be vanished in the state
of thermal equilibrium. Therefore we need the third condition about the deviation 
from thermal equilibrium. The second condition
 is  appropriate for ensuring a different decay rate for particles and antiparticles \cite{mkn}.

The baryon number, C and CP violations can be showed throughout the 
sphaleron rate and the CKM-matrix in models \cite{SME}. The sphaleron 
rate tells us about the baryon number violation and the non-zero phases
 of CKM-matrix tell us about CP violation. 

It is well-known that in order to ensure sufficient the third condition, 
deviations from thermal equilibrium should be large enough, and therefore, 
the EWPT should be the first order phase transition. The EWPT is the transition
between symmetric phase to asymmetric  one  in order to generate masses for particles.
Therefore, the EWPT is related to the mass of the Higgs boson \cite{mkn}.

In the basic model of particles, the first and second conditions can be satisfied,
but conditions on thermal imbalance is difficult to satisfy. So the analysis of
the third condition is the only approach at present in order to explain the baryon asymmetry.

Why is the first order phase transition? The effective potential
is a function of temperature and vacuum expectation values (VEVs).
For very large temperature, it has only one minimum at the zero, and the symmetry is restored.
As temperature goes to $T_0$, the non-zero
second minimum appears, this is the sign of symmetry breaking. When temperature
 at the critical temperature ($T_c<T_0$), the values of the effective potential at 
 two minimums are equal together, the symmetry breaking is turned on. 
 And at the critical temperature, if the two minimums are separated by a
potential barrier, the phase transition will occur with bubble
nucleations. Inside the bubbles,  the scalar field stores a
non-zero expectation value. If the bubble nucleation rate
exceeds the universe's expansion rate,
 the bubbles collide and eventually fill all space \cite{mkn}. Such a transition is called
 the first order phase transition. It is very violent and one can expect large deviations
  from thermal equilibrium \cite{mkn}. The other possible scenario takes place if
the two minimums are never separated by a potential barrier.  In this case, the
phase transition is a smooth transition, not violent or the second order phase
transition.

To study the EWPT, ones consider the hight-temperature effective potential as  follows

\[ V_{eff}=D.(T^2-{T'}^2_0){v}^2-E. Tv^3+\frac{\lambda_T}{4}v^4, \]
where $v$ is the VEV of Higgs. In order to have the first order phase transition,
 the strength of phase transition should be  larger than the unit, i.e., $\frac{v_c}{T_c}\ge 1$.

The EWPT has been investigated in the Standard Model (SM) \cite{mkn, SME} and 
in various extended models \cite{BSM, SMS, dssm, munusm,  majorana, thdm, ESMCO, lr}. 
Also, a very interesting research has showed that dark matter as triggers of the electroweak 
phase transition \cite{elptdm}. For the SM, the strength of  the EWPT is larger than the unit  at 
the electroweak scale, appears too weak
 for the experimentally allowed mass of the SM scalar Higgs boson \cite{mkn,SME}; therefore, 
 it seems that electroweak baryogenesis requires a new physics beyond the 
 SM at weak scale \cite{BSM}.

Before the neutral Higgs was found, the study of phase transitions in most of models 
has been focused on two basic issues: determining the order of phase transition  and the neutral 
 Higgs mass. For the SM, the first order phase transition problem has one variable 
 which is the mass of the neutral Higgs boson. However, for the extended models,
  this problem has at least two variables, the first one is  the Higgs mass and the rest
   includes the masses of heavy particles. Recently, the neutral Higgs has been discovered 
   by the LHC \cite{higgs}, so the electroweak phase transition problem is reduced 
   by one variable. This gives a lot of hope for the extended models in examining the EWPT. 
   The remarkable successes  of the last survey  are:

\begin{itemize}
\item The SM cannot explain this phenomenon: sources of
 CP violation is smaller than Baryon Asymmetry of Universe (BAU)
   and no the first order phase transition with
 the large mass of the neutral Higgs or other word speaking the SM is not enough triggers for the
 first order phase transition to turn on \cite{mkn}.
\item The extended models  such as the two Higgs doublet model (TDHM)
 or MSSM, can explain the  baryon asymmetry,  because sources
 of CP violation in these models are
stronger than in the SM and they have the first order phase transition with
the mass of neutral Higgs about 120 GeV. Triggers for the first order phase
transition in these models are heavy bosons, or dark matter candidates
\cite{majorana, thdm, ESMCO, elptdm}.
\end{itemize}

Among the extended models, the models
based on $\mathrm{SU}(3)_C\otimes \mathrm{SU}(3)_L \otimes
\mathrm{U}(1)_X$ gauge  group (called 3-3-1 for short)
\cite{ppf,flt}  have
some intriguing features such as they can  answer on the
generation problem \cite{ppf,flt}, provide the electric charge
quantization \cite{chargequan}, etc. 
According to the above summarizing,  we hope that the  3-3-1 models
 can also  answer the problem of baryon
asymmetry in our Universe.

The current 3-3-1 models have many different forms, but the core
is based on the above mentioned gauge group. The most
disadvantage of the 3-3-1 models is the complication in the Higgs
sector, namely these models  need at least three  Higgs triplets to
generate masses of fermions. Recently there are attempts to solve
this problem, and some models with the simplest Higgs
sector (with only  two $\mathrm{SU(3)}_L$ Higgs triplets)
have been constructed.

With such  a group structure, the 3-3-1 models  must have at least
two Higgs triplets \cite{ecn331,rm331}. Therefore, the number of
bosons in the 3-3-1 models will many more than in the SM and
symmetry breaking structure is different to the SM.

In the present work, we consider the EWPT in the reduced minimal 3-3-1 (RM331) model
 \cite{rm331} because of its simplicity. This model has the minimal  leptonic content
 (with only the SM leptons) and the
presence of bileptons: a singly and a doubly charged gauge bosons,
$V^{\pm}$ and $U^{\pm\pm}$,  the
 heavy neutral boson $Z_2$ and the exotic quarks.
 This model also has two Higgs triplets. Therefore, the physical scalar
 spectrum of the RM331 model is composed by a doubly charged
 scalar $h^{++}$ and two neutral scalars $h_1$ and $h_2$ \cite{rm331}.
 These new particles and exotic quarks can be triggers for the first order phase transition.

 The plan of the paper is as follows. In
Sec.\ref{sec2} we give a review of the RM331 about the boson,
lepton and Higgs sectors. In Sec. \ref{sec3}, we find the
effective potential in the RM331 that has the contribution from
heavy bosons and a contribution part like in the  SM. In Sec.
\ref{sec4}, we calculate in details structure of phase transition
in the RM331, find the first order phase transition and show
constraints on mass of  charged Higgs boson. Finally, we
summarize  and make outlooks in Sec. \ref{sec5}.

\section{A review of the RM331 model}
\label{sec2}

The fermion content of the RM331 model is the same as of the
minimal 3-3-1 model \cite{ppf}. The difference is only in the
Higgs sector.

\subsection{Higgs potential}

The Higgs potential in the RM331 \cite{rm331} is given by
 \bea V(\chi,\rho)&=&\mu^2_1\rho^\dagger\rho+
\mu^2_2\chi^\dagger\chi+\la_1(\rho^\dagger\rho)^2+\la_2(\chi^\dagger\chi)^2
\crn & &\mbox{}
+\la_3(\rho^\dagger\rho)(\chi^\dagger\chi)+\la_4(\rho^\dagger\chi)(\chi^\dagger\rho)\,.
\label{potential} \eea

The scalar sector contains only two Higgs scalar triplets
\cite{rm331} \be
\rho = \left(\begin{array}{c}\rho^+ \\ \rho^0 \\
\rho^{++}\end{array}\right) \sim \left({\bf 3}, 1\right), \hs
\chi = \left(\begin{array}{c}\chi^- \\ \chi^{--} \\
\chi^0\end{array}\right) \sim \left({\bf 3}, -1\right).
\label{higg}\ee

 Expansion of  $\rho^0$ and $\chi^0$  around
their VEVs is usually \be \rho^0 \,,\, \chi^0 \rightarrow
\frac{1}{\sqrt{2}}(v_{\rho\,,\,\chi}+ R_{\rho\,,\,\chi}
+iI_{\rho\,,\,\chi}). \label{chankhong} \ee

This potential immediately gives us two charged Goldstone bosons
$\rho^\pm$ and $\chi^\pm$  which are eaten by the gauge bosons
$W^\pm$ and $V^\pm$.

Let us resume content of the Higgs sector: the physical scalar
spectrum of the RM331 model is composed by a doubly charged scalar
$h^{++}$ and two neutral scalars $h_1$  and $h_2$. Since the
lightest neutral field, $h_1$, is basically a $SU(2)_L$ component,
we identify it as the SM  Higgs boson. In the effective limit
$v_\chi \gg v_\rho$, the Higgs content can be summarized as
follows \be
\rho = \left(\begin{array}{c}G_{W^+} \\ \fr{v_\rho}{\sqrt{2}}
+ \fr{1}{\sqrt{2}}(h_1 + i G_Z)\\
h^{++}\end{array}\right), \hs
\chi = \left(\begin{array}{c}G_{V^-} \\ G_{U^{--}} \\
\fr{v_\chi}{\sqrt{2}} + \fr{1}{\sqrt{2}}(h_2 + i
G_{Z^\prime})\end{array}\right) \label{effHiggs},\ee where the
Higgs masses are given by: \bea M^2_{h_1} & = &\left(\la_1
-\frac{\la^2_3}{4\la_2}\right)v^2_\rho,\hs M^2_{h_2}=\la_2v^2_\chi
+\frac{\la^2_3}{4\la_2}v^2_\rho, \label{klhiggs}\\
M^2_{h^{--}} & = &\frac{\lambda_4}{2}(v^2_\chi + v^2_\rho). \eea

\subsection{Gauge boson sector}\label{sectgauge}

The masses of gauge bosons  appear in the Lagrangian part
\be
\mathcal{L}=\left( \mathcal{D}_{\mu }\chi \right) ^{\dagger }\left( \mathcal{%
D}^{\mu }\chi \right) +\left( \mathcal{D}_{\mu }\rho \right)
^{\dagger }\left( \mathcal{D}^{\mu }\rho \right) , \label{deriv}
\ee where  \be \mathcal{D}_\mu = \partial_\mu -ig A^a_\mu
\frac{\la ^a}{2}-ig_X X \fr{\la_9}{2} B_\mu, \label{dhhp} \ee with
$\la_9 =\sqrt{\fr 2 3}\, \textrm{diag}(1,1,1)$ so
$\textrm{Tr}(\la_9 \la_9)=2$. The couplings of $SU(3)_L$ and
$U(1)_X$ satisfy the relation:
 \[ \fr{g_X^2}{g^2}= \fr{6 s_W^2}{1-4 s_W^2},\]
where $c_W=\cos\theta_W$, $s_W=\sin\theta_W$, $t_W=\tan \theta_W$
with $\theta_W$ is the Weinberg  angle.

Substitution of the expansion in  Eq. (\ref{chankhong}) into
(\ref{deriv}) leads to the following result
 \bea && W^{\pm}=\frac{A^1 \mp
iA^2}{\sqrt{2}}\hs \rightarrow \hs m_{W^{\pm }}^{2}
 =\frac{g^{2}v_{\rho }^{2}}{4},\crn
&& V^{\pm}=\frac{A^4 \pm iA^5}{\sqrt{2}} \hs \rightarrow \hs
m_{V^{\pm }}^{2}
 =\frac{g^{2}v_{\chi }^{2}}{4},\label{klw} \\
&&
 U^{\pm \pm}=\frac{A^6 \pm iA^7}{\sqrt{2}} \hs \rightarrow \hs m_{U^{\pm \pm }}^{2}
 =\frac{g^{2}\left( v_{\rho }^{2}+v_{\chi }^{2}\right) }{4}\nn
\eea

From (\ref{klw}), it follows that  $v_\rho = 246$ GeV and  the
relation
 \[ m^2_U-m^2_V=m^2_W.\]

In the neutral gauge boson sector, with  the basis ($A^3_{ \mu},
A^8_{\mu}, B_\mu$),  mass  matrix  is given by
 \[ M^2 = \fr{g^2}{4}\left(%
\begin{array}{ccc}
  v^2_\rho & -\fr{ v^2_\rho}{\sqrt{3}} & -2 \kappa v_\rho^2 \\
  -\fr{ v^2_\rho}{\sqrt{3}} & \fr 1 3 (v^2_\rho + 4 v^2_\chi) &
  \fr{2}{\sqrt{3}}(v^2_\rho + 2 v^2_\chi) \\
  -2 \kappa v_\rho^2 & \fr{2}{\sqrt{3}}(v^2_\rho + 2 v^2_\chi) &
  4 \kappa^2 (v^2_\rho +  v^2_\chi), \\
\end{array}%
\right), \] where $\kappa=\fr{g_X}{g}$. We can easily identify the
photon field $A_\mu$ as well as the massive neutral $Z$ and
$Z^\prime$ bosons \cite{dng} \bea A_\mu &=& s_W A^3_\mu +
c_W(\sqrt{3} t_W A^8_\mu +\sqrt{1-3t^2_W} B_\mu),\crn
 Z_\mu &=& c_W A^3_\mu -
s_W(\sqrt{3} t_W A^8_\mu +\sqrt{1-3t^2_W} B_\mu), \nn \eea and \[
Z^\prime_\mu = -\sqrt{1-3t^2_W} A^8_\mu + \sqrt{3} t_W B_\mu \]
where the mass-square matrix for $\{Z, \:  Z^\prime\}$ is given
by \[ \left(%
\begin{array}{cc}
m^2_Z & m^2_{ZZ'} \\
m^2_{ZZ'} & m^2_{Z'} \\
\end{array}%
\right)\] with \bea m^2_Z &=& \frac{1}{4}
\frac{g^2}{\cos^2\theta_W}  v_{\rho}^2,\crn m^2_{Z'} &=&
\frac{1}{3} \: g^2 \left[\frac{\cos^2\theta_W}{1\!-\!4
\sin^2\theta_W} \: v_{\chi}^2 + \:  \frac{1\!- \!4
\sin^2\theta_W}{4\!\cos^2\theta_W}\:v_{\rho}^2\right], \crn
m^2_{ZZ'} &=& \frac{1}{4\sqrt{3}} \, g^2
 \frac{\sqrt{1\!-\!4 \sin^2\theta_W}}{\cos^2\theta_W} v_{\rho}^2.
\nn \eea Diagonalizing the mass matrix gives the mass eigenstates
$Z_1$ and $Z_2$
 which can be taken as mixtures
\bea
Z_1 & = & Z\cos \phi  - Z^\prime \sin \phi, \crn
  Z_2 & = & Z \sin \phi + Z^\prime \cos \phi.
\nn  \eea The mixing angle $\phi$ is given by
 \[  \tan  2\phi = \fr{m^2_{ Z} -  m^2_{ Z_ 1} }{
m^2_{ Z_2 } -  m^2_{ Z}},\] where $m_{Z_1}$ and $m_{Z_2}$ are the
{\it physical} mass eigenvalues \bea
m^2_{Z_1}&=&\frac{1}{2}\left\{m_{Z'}^2+m_{Z}^2-[(m_{Z'}^2-m_{Z}^2)^2-
4(m_{ZZ'}^2)^2]^{1/2}\right\},\crn
m^2_{Z_2}&=&\frac{1}{2}\left\{m_{Z'}^2+m_{Z}^2+[(m_{Z'}^2-m_{Z}^2)^2-
4(m_{ZZ'}^2)^2]^{1/2}\right\}.\nn \eea In  diagonalization, the
mass of $Z_1$ is approximately proportional to $v_{\rho}$ (because
$v_{\chi} \gg v_{\rho}$), so $Z_1$ likes the  neutral   gauge boson $Z$
 in the SM. The mass of the new heavy boson $Z_2$ depends on $v_{\rho}$ and
$v_{\chi}$;   in addition, $m_W=80.39$ GeV and $v_{\rho_0}=246$
GeV \cite{pdg}. Choosing $v_{\chi_0}=4$ TeV \cite{dkck, rm331}, we
obtain $m_V=1307.15$ GeV and $m_U=1309.62$ GeV. If we choose
 $s^2_W=0.23116$ \cite{pdg}, we derive $m_{Z_1}\approx m_{Z}=91.68 $
  GeV and $m_{Z_2}=4.821$ TeV. Hence we can
 approximate $m_{Z_2}\approx 1.2 v_{\chi}$.

\subsection{Fermion sector}

The fermion sector in the model under consideration is the same as
in the minimal 3-3-1 model \cite{ppf}. The Yukawa couplings give
the exotic quark masses \cite{rm331} \bea L^{exot}_{Yuk} & =
&\la^T_{11}\bar Q_{1L}\chi T_{R} + \la^D_{ij}\bar Q_{iL}\chi^*
D_{jR} + H.c.\crn & = &\la^T_{11} (\bar{u}_{1L} \chi^- +
\bar{d}_{1L} \chi^{--} + \bar{T}_{ L} \chi^0) T_{ R} \crn & &+
\la^D_{ij} (\bar{d}_{i L} \chi^+ - \bar{u}_{i L} \chi^{++} +
\bar{D}_{i L} \chi^{0*}) D_{j R}+ H.c. \label{exoticquarks} \eea
When the $\chi$ field develops its VEV, these couplings lead to
the mass matrix in the basis $(T\,,\,D_2\,,\,D_3)$ \[ M_J=
\fr{v_\chi}{\sqrt{2}}
\left(%
\begin{array}{ccc}
\la^T_{11} & 0 & 0 \\
0& \la^D_{22} & \la^D_{23}\\
0 & \la^D_{32} & \la^D_{33}.\\
\end{array}%
\right) \] So the exotic quarks have masses around few TeVs because their
 masses are proportional to $v_\chi$. Therefore we  see that masses 
 of exotic quarks approximately equal to $m_{Z_2}$,  but they only involve in the transition 
 phase $SU(3) \rightarrow SU(2)$. 

For usual  quarks, the Yukawa couplings give the masses of them
through the triplet $\rho$. Therefore, also like the SM, the usual
quarks only involve in the transition phase $SU(2) \rightarrow U(1)$.

However, the charged lepton masses arise from the effective
dimension five operator through the couplings of  both $\chi$ and
$\rho$ with the following Lagrangian \cite{rm331}
 \be L^l_{Yuk}
=\frac{\kappa_l}{\La}\left(\overline{f^c_L}\rho^*\right)\left(\chi^\dagger
f_L \right) + H.c.\label{lly} \ee

From Lagrangian (\ref{lly}), we  obtain $ m_l=
\fr{v_\chi}{\La}\kappa_l v_\rho$, the coupling constant
$v_{\chi}/\Lambda \approx 1$, so $ m_l \approx \kappa_l v_\rho $.
Finally the masses of charged leptons  depend only on $v_\rho$.
Therefore, they only involve
 in the transition phase $SU(2) \rightarrow U(1)$. Taking into account $m_e = 0.5$ MeV,
  $m_{\mu} = 105$ MeV, $m_{\tau} = 1.77$ GeV, ones get $k_e = 2 \times 10^{-5}$,
  $k_\mu = 4.3 \times 10^{-3}$ and $k_\tau = 7.2 \times 10^{-2}$.

\section{Effective potential in RM331}
\label{sec3} 
From the Higgs potential we  obtain $V_{0}$ 
depending on VEVs as  follows \[ V_{0}(v_{\chi},
v_{\rho})=\mu^2_1v_{\chi}^2+\mu^2_2v_{\rho}^2
+\lambda_1v_{\chi}^4+\lambda_2v_{\rho}^4+\lambda_3v_{\chi}^2v_{\rho}^2.\] 
Here $V_{0}$ has quartic form
like in the SM, but it depends on two variables $v_{\chi}$ and
$v_{\rho}$ and has the mixing between $v_{\chi}$ and $v_{\rho}$.
 However,  developing  the potential (\ref{potential}), we obtain two minimum  equations.
 Therefore, we can transform the mixing between $v_{\chi}$ and $v_{\rho}$ to
 the form that  depends only on $v_{\chi}$ or $v_{\rho}$. Hence,
 we can write  $V_0(v_{\chi}, v_{\rho})=V_0(v_\chi)+V_0(v_{\rho})$.

In order to derive effective potential, starting from the Higgs Lagrangian 
 and using the principle of least action,  we arrive at the equation of motion for fields. 
  Expanding  Higgs fields around VEVs and averaging over  space for all fields, 
 we  obtain the one-loop effective potential (for details, see  Ref.\cite{mkn}).
 
The full  Higgs Lagrangian  in the RM331 model is given by 
\[ \mathcal{L}=\left( \mathcal{D}_{\mu }\chi \right) ^{\dagger }\left( \mathcal{%
D}^{\mu }\chi \right) +\left( \mathcal{D}_{\mu }\rho \right)
^{\dagger }\left( \mathcal{D}^{\mu }\rho \right)+V(\chi,\rho), \]
where
\bea V(\chi,\rho)&=&\mu^2_1\rho^\dagger\rho+
\mu^2_2\chi^\dagger\chi+\la_1(\rho^\dagger\rho)^2+\la_2(\chi^\dagger\chi)^2
\crn & &\mbox{}
+\la_3(\rho^\dagger\rho)(\chi^\dagger\chi)+\la_4(\rho^\dagger\chi)(\chi^\dagger\rho).\nn
\eea

Expanding  $\rho$ and $\chi$ around  $v_{\rho}$ and $v_{\chi}$ which  are considered as
 variables \footnote{At $0^K$, $v_{\rho}\equiv v_{\rho_0}=246$ GeV and 
 $v_{\chi}\equiv v_{\chi_0}=4 \div 5$ TeV. In this work, we choose $v_{\chi_0}=4$ TeV }, so we obtain
\[
\mathcal{L}=\frac{1}{2}\partial^{\mu}v_\chi\partial_{\mu}v_\chi+
\frac{1}{2}\partial^{\mu}v_\rho\partial_{\mu}v_\rho+V_0(v_{\chi}, v_{\rho})
+\sum_{boson}m^2_{boson}(v_{\chi}, v_{\rho})W^{\mu}W_{\mu},
\]
where $W$ runs over all gauge fields and Higgs bosons. Through the boson mass
 formulations as in the above sections, we can split masses of particles into two
 parts as follows 
 \[ m^2_{boson}(v_{\chi}, v_{\rho})=
 m^2_{boson}(v_{\chi})+m^2_{boson}(v_{\rho}).\] 

The effective potential is the function that depends on VEVs and
temperature. The masses of particles depend on VEV of Higgs bosons.
Therefore, when we consider the effective potential, we must
consider contributions from  fermions and bosons. However,  for fermions, 
we have retained 
here only the top quark and exotic quarks, which dominate over the contributions 
from the other fermions \cite{mkn}. And in the  RM331 model, there are two 
VEVS so we have two motion equations according  to  $v_\chi$ and $v_{\rho}$
\bea
 \partial^{\mu}v_{\chi}\partial_{\mu}v_{\chi}+\frac{\partial V_0(v_{\chi})}{\partial 
v_\chi}+\sum \frac{\partial m^2_{bosons}(v_{\chi})}{\partial v_\chi}W^{\mu}
W_{\mu}+\sum\frac{\partial m_{exotic-quarks}(v_{\chi})}{\partial v_\chi}Q\bar{Q} &= &0,\label{mchi}\\
\partial^{\mu}v_{\rho}\partial_{\mu}v_{\rho}+\frac{\partial V_0(v_{\rho})}{\partial v_\rho}
+\sum \frac{\partial m^2_{bosons}(v_{\rho})}{\partial v_\rho}W^{\mu}W_{\mu}+
\frac{\partial m_{top-quark}(v_{\rho})}{\partial v_\rho}t\bar{t}&=&0. \label{mrho}
\eea

The RM331 has the gauge bosons: two massive like the SM
bosons $Z_1$ and  $W^{\pm}$
 and the new  heavy neutral boson $Z_2$, the singly and doubly charged gauge
  bosons $U^{\pm\pm}$ and $V^{\pm}$; and
 two doubly charged Higgses $h^{++}$ and  $h^{--}$, one heavy neutral Higgs $h_2$ and
 one like-SM Higgs $h_1$.  Masses of gauge bosons and Higgses 
 in the RM 331 model are  presented  in Table \ref{tab1}
 
 \begin{table}
\caption{Mass formulations of bosons in the RM331 model}
\bc
\begin{tabular}{|l|l|c|c|}
\hline
 Bosons&$m^2(v_\chi,v_\rho)$ & $ m^2(v_\chi)$  & $m^2(v_\rho)$    \\
\hline
$m_{W^{\pm }}^{2}$&$\frac{g^{2}v_{\rho }^{2}}{4}$ & 0 &$80.39^2$ $(\mathrm{GeV})^2$ \\
\hline
$m_{V^{\pm }}^{2}  $&$\frac{g^{2}v_{\chi }^{2}}{4}$ &$1307.15^2$ $(\mathrm{GeV})^2$ &0\\
\hline
$m_{U^{\pm \pm }}^{2} $&$\frac{g^{2}\left( v_{\rho }^{2}+
v_{\chi }^{2}\right) }{4}$&$1307.15^2$ $(\mathrm{GeV})^2$ &$80.39^2$ $(\mathrm{GeV})^2$ \\
\hline
$m^2_{Z_1}\sim m^2_{Z} $ &$\frac{1}{4}\frac{g^2}{\cos^2\theta_W} 
 v_{\rho}^2$ &0 &$91.682^2$ $(\mathrm{GeV})^2$ \\
\hline
$m^2_{Z_2}\sim m^2_{Z'}$ &$\frac{1}{3} \: g^2 \left[\frac{\cos^2\theta_W}{1\!-\!4
\sin^2\theta_W} \: v_{\chi}^2 + \:  \frac{1\!- \!4
\sin^2\theta_W}{4\!\cos^2\theta_W}\:v_{\rho}^2\right]$ &$4.8^2$ 
$(\mathrm{TeV})^2$ & $14.53^2$ $(\mathrm{GeV})^2$ \\
\hline
$m^2_{h_1} $&$ \left(\la_1 -\frac{\la^2_3}{4\la_2}\right)v^2_\rho$ &0&
$125^2$ $(\mathrm{GeV})^2$\\
\hline
$m^2_{h^{--}} $&$ \frac{\lambda_4}{2}(v^2_\chi + v^2_\rho)$ 
&$\frac{\lambda_4}{2}v^2_\chi$ &$\frac{\lambda_4}{2} v^2_\rho$\\
\hline
$m^2_{h_2}$&$\la_2v^2_\chi +\frac{\la^2_3}{4\la_2}v^2_\rho$ &
$\la_2v^2_\chi $ &$\frac{\la^2_3}{4\la_2}v^2_\rho$ \\
\hline
\end{tabular}
\ec
\label{tab1}
\end{table}

From two equations (\ref{mchi}) and  (\ref{mrho}) and averaging over space by 
using Bose-Einstein and Fermi-Dirac distributions for bosons and fermions, we obtain 
the following effective potentials
\bea V_{eff}(v_{\chi})&= &V_{0}(v_{\chi})+\frac{3}{64\pi^2}
\left(m^4_{Z_2}(v_{\chi})\ln\frac{m^2_{Z_2}(v_{\chi})}{Q^2}-12m^4_Q(v_{\chi})
\ln\frac{m^2_Q(v_{\chi})}{Q^2}\right)\crn
&&+\frac{1}{64\pi^2}\left(m^4_{h_2}(v_{\chi})\ln\frac{m^2_{h_2}(v_{\chi})}{Q^2}+2m^4_{h^{++}}(v_{\chi})
\ln\frac{m^2_{h^{++}}(v_{\chi})}{Q^2}\right)\crn
&&+\frac{3}{64\pi^2}\left(2m^4_U(v_{\chi})\ln\frac{m^2_U(v_{\chi})}{Q^2}+2m^4_V(v_{\chi})
\ln\frac{m^2_V(v_{\chi})}{Q^2}\right)\crn
&&+\frac{T^4}{4\pi^2}\left[F_{-}\left(\frac{m_{h_2}(v_{\chi})}{T}\right)+
2F_{-}\left(\frac{m_{h^{++}}(v_{\chi})}{T}\right)+12F_{+}\left(\frac{m_Q(v_{\chi})}{T}\right)\right]\crn
&&+\frac{3T^4}{4\pi^2}\left[F_{-}\left(\frac{m_{Z_2}(v_{\chi})}{T}\right)
+2F_{-}\left(\frac{m_U(v_{\chi})}{T}\right) +2F_{-}\left(\frac{m_V(v_{\chi})}{T}\right)\right], \nn\eea 
and
\bea V_{eff}(v_{\rho})&= &V_{0}(v_{\rho})+\frac{3}{64\pi^2}
\left(m^4_{Z_1}(v_{\rho})\ln\frac{m^2_{Z_1}(v_{\rho})}{Q^2}+
m^4_{Z_2}(v_{\rho})\ln\frac{m^2_{Z_2}(v_{\rho})}{Q^2}
\right. \crn &&+ \left.2m^4_W(v_{\rho})\ln\frac{m^2_W(v_{\rho})}{Q^2}
+2m^4_U(v_{\rho})\ln\frac{m^2_U(v_{\rho})}{Q^2}-4m^4_t(v_{\rho})
\ln\frac{m^2_t(v_{\rho})}{Q^2}\right)\crn
&&+\frac{1}{64\pi^2}\left(m^4_{h_1}(v_{\rho})\ln\frac{m^2_{h_1}(v_{\rho})}{Q^2}-
m^4_{h_2}(v_{\rho})\ln\frac{m^2_{h_2}(v_{\rho})}{Q^2}+2m^4_{h^{++}}
\ln\frac{m^2_{h^{++}}(v_{\rho})}{Q^2}\right)\crn
&&+\frac{T^4}{4\pi^2}\left[F_{-}\left(\frac{m_{h_1}(v_{\rho})}{T}\right)-
F_{-}\left(\frac{m_{h_2}(v_{\rho})}{T}\right)
+2F_{-}\left(\frac{m_{h^{++}}(v_{\rho})}{T}\right)\right]\crn
&&+\frac{3T^4}{4\pi^2}\left[4F_{+}\left(\frac{m_t(v_{\rho})}{T}\right)
+F_{-}\left(\frac{m_{Z_1}(v_{\rho})}{T}\right)+F_{-}
\left(\frac{m_{Z_2}(v_{\rho})}{T}\right)\right.\crn &&+ \left.
2F_{-}\left(\frac{m_W(v_{\rho})}{T}\right)+2F_{-}
\left(\frac{m_U(v_{\rho})}{T}\right)\right], \nn\eea 
where \bea
F_{\mp}\left(\frac{m_{\phi}}{T}\right)&=&\int^{\frac{m_{\phi}}{T}}_0\alpha
J^{(1)}_{\mp}(\alpha,0)d\alpha, \crn
J^{(1)}_{\mp}(\alpha,0)&=&2\int^{\infty}_{\alpha}\frac{(x^2-\alpha^2)^{1/2}}{e^{x}\mp
1}dx. \nn\eea

The total effective potential in the  RM331 model can be rewritten as follows
\[ V^{RM331}_{eff}=V_{eff}(v_\chi)+V_{eff}(v_\rho).\]

\section{Electroweak phase transition}

\label{sec4} The symmetry breaking in the RM331 can take place
sequentially: due to two scales of symmetry breaking which  are very
different, $v_{\chi_0} \gg  v_{\rho_0}$ ($v_{\chi_0}\sim 4-5$ TeV \cite{dkck, rm331},
 $v_{\rho_0}=246$ GeV) and because of the accelerating universe.
  The symmetry breaking  $SU(3)\rightarrow SU(2)$ takes place
   before the symmetry breaking $SU(2)\rightarrow U(1)$.

The symmetry breaking $SU(3)\rightarrow SU(2)$ through $\chi_0$, generates
   masses of the heavy gauge bosons such  as
   $U^{\pm\pm}$, $V^{\pm}$, $Z_2$, and the exotic quarks.
   Therefore, the phase transition $SU(3)\rightarrow SU(2)$ only depends on $v_{\chi}$.
   When our universe has been expanding and cooling due to $v_{\rho_0}$ scale,
   the symmetry breaking or phase transition $SU(2)\rightarrow U(1)$ is turned on
   through $\rho_0$, which generates  masses of the  SM particles and the last  part of
 masses of   $U^{\pm\pm}$. Therefore phase transition $SU(2)\rightarrow U(1)$
 only depends on $v_{\rho}$.
For the current universe having the baryon asymmetry, this
asymmetry must exist from the initial
 conditions of the universe. If the early universe has not the baryon asymmetry, the current
 universe will not have it \cite{mkn}. In other words, this asymmetry exists throughout periods of the
  universe to date. In order to describe baryogenesis,
   models must satisfy three conditions
  given by Sakharov. Especially in the third condition, the spontaneous symmetry breaking
   must be associated with the first order phase transition.
   In the RM331 model, the spontaneous
 symmetry breaking takes place in two different energy
  scales or other word speaking the electroweak phase
    transition is the combination of two different phase transitions. Therefore, if the RM331
  model well describes  this phenomenon, both two phase transitions are the first order
 phase transition. In contrast, if one of two phase transitions is not the first order phase
  transition, the RM331 model will not fully describe this asymmetry, since this model
  does not ensure the continuity of baryogenesis in the universe.

Through the  boson mass formulations in the above sections, we see that boson $V^{\pm}$
 only involves in the first phase transition $SU(3) \rightarrow SU(2)$. The gauge bosons
 $Z_1$, $W^{\pm}$ and $h_1$ only involve in the second
  phase transition $SU(2) \rightarrow U(1)$. However,
 $U^{\pm\pm}$, $Z_2$ and $h^{--}$ involve in both two phase transitions.

With that structure of phase transition, we see that the mass of
$U^{\pm\pm}$, is generated by both two phase transitions. When the
universe at $v_{\chi_0}$ scale, the symmetry breaking
$SU(3)\rightarrow SU(2)$ generates masses for the exotic quarks
and a part of $U^{\pm\pm}$, or other word speaking,  it  eaten one
Goldstone boson $\chi^{\pm\pm}$ of triplet $\chi$. When the
universe cool to $v_{\rho_0}$ scale, the symmetry breaking
$SU(2)\rightarrow U(1)$ is turned on, it generates masses for the
SM particles and the last part of $U^{\pm\pm}$, or $U^{\pm\pm}$
eaten the other Goldstone boson $\rho^{\pm\pm}$ of triplet $\rho$.

\subsection{Phase transition $SU(3) \rightarrow SU(2)$}

This phase transition involves exotic quarks, heavy bosons, without 
involvement of the SM particles, the effective potential of the
 EWPT $SU(3) \rightarrow SU(2)$ is $V_{eff}(v_{\chi})$. 
\bea
V_{eff}(v_{\chi})&= &V_{0}(v_{\chi})+\frac{3}{64\pi^2}
\left(m^4_{Z_2}(v_{\chi})\ln\frac{m^2_{Z_2}(v_{\chi})}{Q^2}-
12m^4_Q(v_{\chi})\ln\frac{m^2_Q(v_{\chi})}{Q^2}\right)\crn
&&+\frac{1}{64\pi^2}\left(m^4_{h_2}(v_{\chi})
\ln\frac{m^2_{h_2}(v_{\chi})}{Q^2}+2m^4_{h^{++}}(v_{\chi})
\ln\frac{m^2_{h^{++}}(v_{\chi})}{Q^2}\right)\crn
&&+\frac{3}{64\pi^2}\left(2m^4_U(v_{\chi})
\ln\frac{m^2_U(v_{\chi})}{Q^2}+2m^4_V(v_{\chi})
\ln\frac{m^2_V(v_{\chi})}{Q^2}\right)\crn
&&+\frac{T^4}{4\pi^2}\left[F_{-}\left(\frac{m_{h_2}(v_{\chi})}{T}\right)
+2F_{-}\left(\frac{m_{h^{++}}(v_{\chi})}{T}\right)+12F_{+}
\left(\frac{m_Q(v_{\chi})}{T}\right)\right]\crn
&&+\frac{3T^4}{4\pi^2}\left[F_{-}\left(\frac{m_{Z_2}(v_{\chi})}{T}\right)
+2F_{-}\left(\frac{m_U(v_{\chi})}{T}\right) +2F_{-}
\left(\frac{m_V(v_{\chi})}{T}\right)\right].\nn
\eea

The symmetry breaking scale is $v_{\chi_0}$  chosen  to be  $4$ TeV 
\cite{dkck, rm331} and the masses of three
exotic quarks are $m_Q$. Therefore, the effective potential can be rewritten as follows
 \[ V_{SU(3) \rightarrow SU(2)}^{eff}=D'(T^2-{T'}^2_0){v_{\chi}}^2-E'
Tv_{\chi}^3 +\frac{\lambda'_T}{4}v_{\chi}^4
\] The minimum  conditions are \[ V_{eff}(v_{\chi_0})=0;\hs
\frac{\partial V_{eff}(v_{\chi})}{\partial v_\chi}(v_{\chi_0})=0;
\hs  \frac{\partial^2 V_{eff}(v_{\chi})}{\partial v_\chi^2}(v_{\chi_0})
=m^2_{h_2}(v_{\chi})|_{v_{\chi}=v_{\chi_0}}, \] where
\bea
D'&=& \frac{1}{24 {v_{\chi_0}}^2} \left\{6 m_U^2(v_{\chi})+ 3m_{Z_2}^2(v_{\chi})+6 m_{V}^2(v_{\chi})
+ 18m_Q^2(v_{\chi})+ 2 m_{h_2}^2(v_{\chi})+2 m_{h^\pm}^2(v_{\chi}) \right\},\crn
{T'}_0^2 &=&  \frac{1}{D}\left\{\frac{1}{4} m_{h_2}^2(v_{\chi}) -
 \frac{1}{32\pi^2v_{\chi_0}^2} \left(6 m_U^4(v_{\chi})+ 3 m_{Z_2}^4(v_{\chi})+6 m_{V}^4(v_{\chi}) -
 36 m_Q^4(v_{\chi}) \right.\right.\crn
&&\qquad \left.\left.+m_{h_2}^4(v_{\chi})+ 2 m_{h^\pm}^4(v_{\chi})\right)\right\},\crn
E' &=& \frac{1}{12 \pi v_{\chi_0}^3} (6 m_U^3(v_{\chi}) + 3 m_{Z_2}^3(v_{\chi}) +
6 m_{V}^3(v_{\chi}) +m_{h_2}^3(v_{\chi})+ 2 m_{h^\pm}^3(v_{\chi})),\crn
\lambda'_T &=&
 \frac{m_{h_2}^2(v_{\chi})}{2 v_{\chi_0}^2} \left\{ 1 - \frac{1}{8\pi^2 v_{\chi_0}^2 m_{h_2}^2(v_{\chi})}
 \left[6 m_V^4(v_{\chi}) \ln \frac{m_V^2(v_{\chi})}{b T^2} +
 3 m_{Z_2}^4(v_{\chi}) \ln \frac{m_{Z_2}^2(v_{\chi})}{b T^2}
 \right.\right.\crn
&&\qquad \left.\left. +6 m_U^4(v_{\chi}) \ln \frac{m_U^2(v_{\chi})}{bT^2}-36 m_Q^4(v_{\chi}) \ln
\frac{m_Q^2(v_{\chi})}{b_F T^2} +m_{h_2}^4(v_{\chi}) \ln \frac{m_{h_2}^2(v_{\chi})}{b
T^2}\right.\right.\crn
&&\qquad \left.\left.+ 2  m_{h^\pm}^4(v_{\chi}) \ln \frac{m_{h^\pm}^2(v_{\chi})}{b
T^2} \right] \right\}.\nn \eea The values of $V_{eff}(v_{\chi})$ at
the two minima become equal at the critical temperature  \be
T'_c=\frac{T'_0}{\sqrt{1-E'^2/D'\lambda'_{T'_c}}}.\label{th} \ee

The problem here is that there are three variables: the masses 
of $h_2$, $h^{--}$ and $Q$. However, for simplicity, following  the ansazt
 in \cite{ESMCO}, we assume $m_{h_2}=X$, $m_{h^{--}}=m_{Q}=K$
  and $m_{Z2}(v_{\chi})=4.821$ TeV.
 Note that the contributions from $h_2, h^{--}, Z_2$,  in this
  phase transition, are $X$ or $K$, different to  their contributions in
  the phase transition $SU(2)\rightarrow U(1)$.
  In order to have the first phase transition, the phase transition strength  must be larger than
  the unit, i.e., $\frac{v_{\chi_c}}{T'_c} \ge 1$.

\begin{figure}[h]
\centering
\includegraphics[height=9cm,width=16cm]{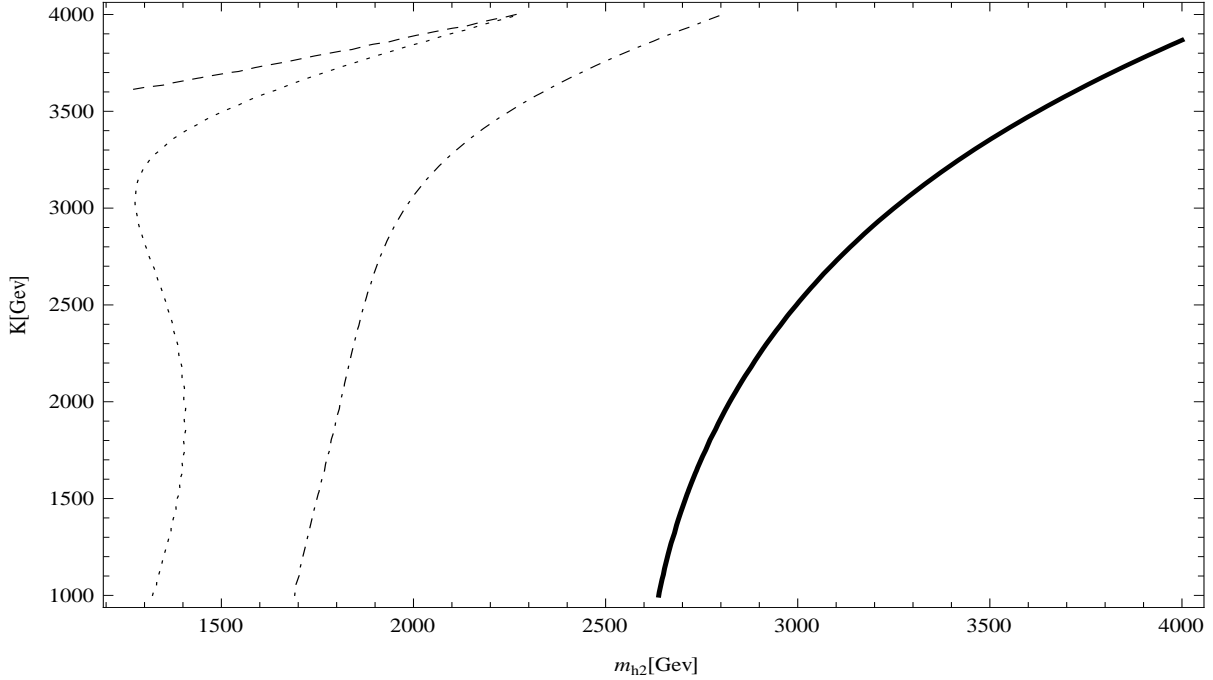}
\caption{When $X > 1$ TeV, the solid contours of $2E'/\lambda'_{T'_c}=1$, the
dashed-dotted contour: $2E'/\lambda'_{T'_c}=2$, the dotted contour: $2E'/\lambda'_{T'_c}=3$,
the dashed contour: $2E'/\lambda'_{T'_c}=5$.}\label{fig:00}
\end{figure}

In figure \ref{fig:00}, we have plotted $K$ as a function of $m_{h_2}(v_{\chi})$, 
with $m_{h_2}(v_\chi)> 1$ TeV.

According to Fig. \ref{fig:00}, if $X$ is larger than $1$ TeV,  the heavy 
particle masses are in range of few TeVs in order to have the first order phase transition.
In addition, this phase transition can be  strong first order.

\subsection{Phase transition $SU(2) \rightarrow U(1)$}

This phase transition dose not involve the exotic quarks and boson 
$V^{\pm}$ and the contribution from  $U^{\mp\mp}$ is equal to $ W^{\mp}$. 
The effective potential of the EWPT $SU(2) \rightarrow U(1)$ is $V_{eff}(v_\rho)$:
\bea V_{eff}(v_{\rho})&= &V_{0}(v_{\rho})+\frac{3}{64\pi^2}
\left(m^4_{Z_1}(v_{\rho})\ln\frac{m^2_{Z_1}(v_{\rho})}{Q^2}
+m^4_{Z_2}(v_{\rho})\ln\frac{m^2_{Z_2}(v_{\rho})}{Q^2}
\right. \crn &&+ \left.2m^4_W(v_{\rho})\ln\frac{m^2_W(v_{\rho})}{Q^2}
+2m^4_U(v_{\rho})\ln\frac{m^2_U(v_{\rho})}{Q^2}-4m^4_t(v_{\rho})
\ln\frac{m^2_t(v_{\rho})}{Q^2}\right)\crn
&&+\frac{1}{64\pi^2}\left(m^4_{h_1}(v_{\rho})\ln\frac{m^2_{h_1}(v_{\rho})}{Q^2}
+m^4_{h_2}(v_{\rho})\ln\frac{m^2_{h_2}(v_{\rho})}{Q^2}+2m^4_{h^{++}}
\ln\frac{m^2_{h^{++}}(v_{\rho})}{Q^2}\right)\crn
&&+\frac{T^4}{4\pi^2}\left[F_{-}\left(\frac{m_{h_1}(v_{\rho})}{T}\right)
+F_{-}\left(\frac{m_{h_2}(v_{\rho})}{T}\right)
+2F_{-}\left(\frac{m_{h^{++}}(v_{\rho})}{T}\right)\right]\crn
&&+\frac{3T^4}{4\pi^2}\left[4F_{+}\left(\frac{m_t(v_{\rho})}{T}\right)
+F_{-}\left(\frac{m_{Z_1}(v_{\rho})}{T}\right)+F_{-}
\left(\frac{m_{Z_2}(v_{\rho})}{T}\right)\right.\crn && \qquad \qquad + \left.
2F_{-}\left(\frac{m_W(v_{\rho})}{T}\right)+
2F_{-}\left(\frac{m_U(v_{\rho})}{T}\right)\right]. \nn\eea 

The minimum  conditions are \[ V_{eff}(v_{\rho_0})=0;\hs
\frac{\partial V_{eff}}{\partial v_{\rho}}(v_{\rho_0})=0;\hs 
 \frac{\partial V_{eff}}{\partial v_\rho}(v_{\rho_0})=
 m^2_{h_1}+m^2_{h_2}(v_{\rho})|_{v_\rho=v_{\rho_0}}. \]

From  the above minimum conditions, we see that in this
 EWPT $m^2_{h_2}(v_{\rho})$  plays the role to generate masses 
 of the last heavy particles and $m^2_{h_1}$ generates  masses of the
 SM particles.

With the symmetry breaking scale equal to $Q \equiv $ $ v_{\rho_0}= $ $v_0=246$ GeV,
the high-temperature expansion of this potential has the form
\[ V_{eff}^{RM331}=D(T^2-T^2_0).v^2_{\rho}-ET|v_{\rho}|^3+\frac{\lambda_T}{4}v^4_{\rho},
\]
where
\bea
D &=& \frac{1}{24 {v_0}^2} \left[6 m_W^2(v_{\rho})+6 m_U^2(v_{\rho})
+3m_{Z_1}^2(v_{\rho}) +3 m_{Z_2}^2(v_{\rho}) \right.\crn
&&\quad \left.+ 6 m_t^2 (v_{\rho})+m_{h_1}^2(v_{\rho})+m_{h_2}^2(v_{\rho})
 + 2 m_{h^\pm}^2(v_{\rho}) \right],\crn
T_0^2 &=&  \frac{1}{D}\left\{\frac{1}{4} (m_{h_1}^2(v_{\rho})+m_{h_2}^2(v_{\rho})) -
 \frac{1}{32\pi^2v_0^2} \left(6 m_W^4(v_{\rho})+6 m_U^4(v_{\rho})
 + 3 m_{Z_1}^4(v_{\rho}) \right.\right.\crn
&&\qquad \left.\left.
 +3 m_{Z_2}^4(v_{\rho}) - 12 m_t^4(v_{\rho}) + m_{h_1}^4(v_{\rho})
 +m_{h_2}^4(v_{\rho}) + 2 m_{h^\pm}^4(v_{\rho})\right)
\right\},\crn
E &=& \frac{1}{12 \pi v_0^3} \left(6 m_W^3(v_{\rho})+6 m_U^3(v_{\rho}) 
+ 3 m_{Z_1}^3(v_{\rho}) +3 m_{Z_2}^3(v_{\rho}) \right.\crn
&&\quad \left. +m_{h_1}^3(v_{\rho})+m_{h_2}^3(v_{\rho}) 
+ 2 m_{h^\pm}^3(v_{\rho})\right),\label{trilinear}\\
\lambda_T &=&
 \frac{m_{h_1}^2(v_{\rho})+ m_{h_2}^2(v_{\rho})}{2 v_0^2}
 \left\{ 1
- \frac{1}{8\pi^2 v_0^2 (m_{h_1}^2(v_{\rho})+m_{h_2}^2(v_{\rho}))}
 \left[6 m_W^4(v_{\rho}) \ln \frac{m_W^2(v_{\rho})}{b T^2} \right.\right.\crn
&&\qquad \left.\left. + 3 m_{Z_1}^4(v_{\rho}) \ln\frac{m_{Z_1}^2(v_{\rho})}{b T^2}
+3 m_{Z_2}^4(v_{\rho}) \ln \frac{m_{Z_2}^2(v_{\rho})}{b T^2}+6 m_U^4(v_{\rho})
 \ln \frac{m_U^2(v_{\rho})}{bT^2}\right.\right.\crn
&&\qquad \left.\left.-12 m_t^4(v_{\rho}) \ln\frac{m_t^2(v_{\rho})}{b_F T^2} 
+ m_{h_1}^4(v_{\rho}) \ln \frac{m_{h_1}^2(v_{\rho})}{b T^2} 
+m_{h_2}^4(v_{\rho}) \ln \frac{m_{h_2}^2(v_{\rho})}{b T^2}\right.\right.\crn
&&\qquad  \qquad  \qquad  \qquad  \left.\left. + 2m_{h^\pm}^4(v_{\rho}) 
\ln \frac{m_{h^\pm}^2(v_{\rho})}{b T^2} \right] \right\}\nn \eea

The effective potential has two minimum  points, the first minimum 
at $v_{\rho}=0$ and the second one at
 $v_{\rho_c}=\frac{2ET_c}{\lambda_{T_c}}$. In the limit $E\rightarrow 0$,
 we has the second order phase transition. In order to have the first order
 phase transition, the phase transition strength has to be  larger than the unit, i.e.,
 $\frac{v_{\rho_c}}{T_c}\geq 1$. The critical temperature $T_c$ is given by
\be T_c=\frac{T_0}{\sqrt{1-E^2/D\lambda_{T_c}}}.\label{th}
\ee
The equation (\ref{th}) is  self-consistent with the critical
temperature because $\lambda_{T_c}$ is a function of $T_c$.
According to the LHC, we take  $m_{h_1}=125$ GeV and put $m_{h_2}(v_\rho)=
Z$, $m_{h^{--}}(v_\rho)=Y$ and $m_{Z2}(v_{\rho})=14.53$ GeV.

\begin{figure}[h]
\centering
\includegraphics[height=9cm,width=16cm]{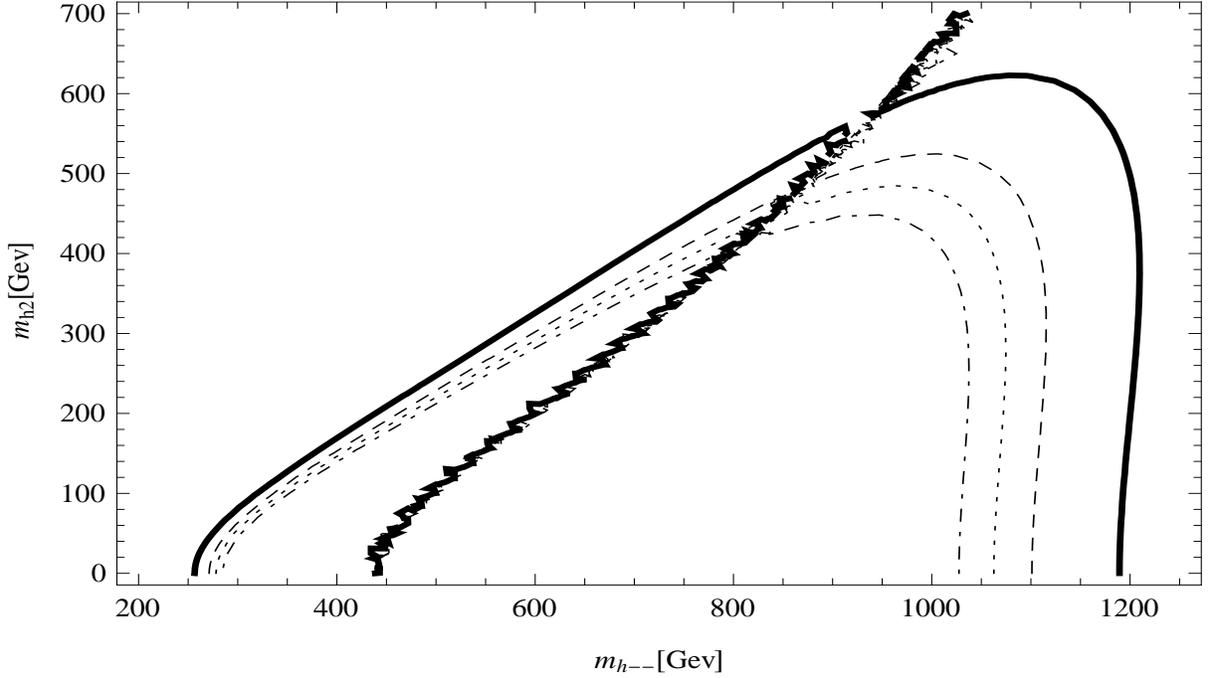}
\caption{When the solid contour of $2E/\lambda_{T_c}=1$, the
dashed contour: $2E/\lambda_{T_c}=1.1$, the dotted contour: $2E/\lambda_{T_c}=1.15$,
the dotted-dashed contour: $2E/\lambda_{T'_c}=1.2$}\label{fig:01}
\end{figure}
In Fig.\ref{fig:01}, we show the masses region of $h_1$ and $h_{++}$ 
where the necessary condition of the first order phase transition is imposed. 
According to Fig.\ref{fig:01}, and by  the numerical evaluation, 
the strength of  the EWPT is in range, 
$1\leq2E/\lambda_{T_c}<5$. Therefore, in the RM331 we always 
have the first order phase transition but it is weak (at the $v_\rho$ scale).
 
The contributions from new particles
(at the first symmetry breaking) make of the first order phase transition
 that the Standard Model cannot. However, there is one thing special, heavy
  particles as $U^{\pm\pm}, h_2, h^ {--}, Z_2$ that contribute only the little part in their total masses.

\begin{figure}[!ht]
\centering
\includegraphics[height=6cm,width=10cm]{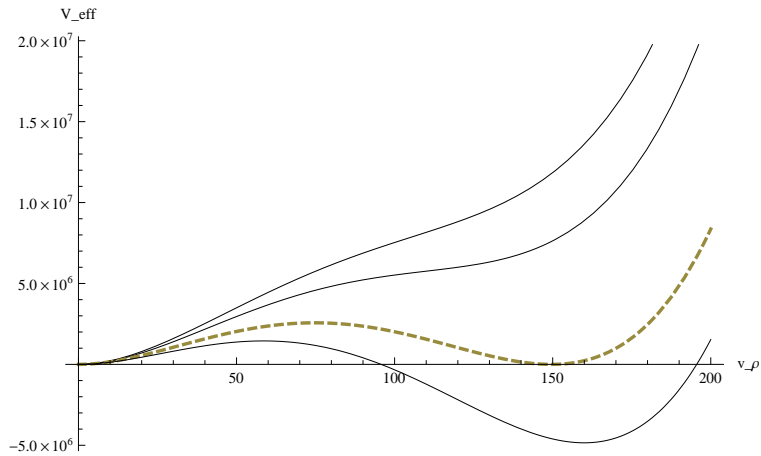}
\caption{For example, the effective potential for $2E/\lambda_{T_c}=1$, with $Y=342$ GeV, $Z=120.109$ GeV. The critical point at $T_c=149.549 \hs Gev$. Solid line: $T=T_c$, lines above solid line: $T>{T_c}$, lines under solid line: $T<T_c$  }\label{fig:02}
\end{figure}

In Fig. \ref{fig:02}, when temperature go close to $ T_c $, the second minimum slowly formed distinct, ie the phase transition nucleation appears. When temperature go to $T_c$, the symmetry breaking phase is turned on and temperature go down $T_c$, the system switches to the symmetry breaking phase.

To finish this section, we conclude  that the effective potential of this model is different
from that of the SM, and  it has the contributions from heavy bosons, which plays
 role of  the triggers for the first order phase transition with $m_{h_1}=125$ GeV.

\subsection{Constraint on mass of charged Higgs boson}
In Section \ref{sec4}, from  the phase transition  $SU(2)\rightarrow U(1)$, we have derived 
\[ 200 \,
\mathrm{GeV}   < Y=m_{h_{++}}(v_\rho) < 1200\,  \mathrm{GeV} \] and 
\[0    <  Z=m_{h_2}(v_\rho) < 624 \,  \mathrm{GeV}.\] 
Therefore, we get
\bea
&& (200 \,\, \mathrm{GeV})^2 < \frac{\lambda_4}{2}v^2_{\rho_0} < (1200\, \mathrm{GeV})^2,
\crn
&&  (0 \,\, \mathrm{GeV})^2  <\frac{\la^2_3}{4\la_2}
v^2_{\rho_0}< (624\, \mathrm{GeV})^2. \label{dk1}
\eea
Taking into account of the recent Higgs boson mass (125 GeV), combining with the
formula (\ref{klhiggs}),  we also obtain
\be \left(\la_1
-\frac{\la^2_3}{4\la_2}\right)v^2_{\rho_0}=  (125 \, \mathrm{GeV})^2.\label{dk2}\ee
Combination of the equations in (\ref{dk1})  and  (\ref{dk2}) leads to
\bea
&& 0 < \lambda_1 < 6.692, \crn
&& 0< \frac{\lambda^2_3}{4\lambda_2} < 6.434, \label{hq1} \crn
&& 1.321< \lambda_4 < 47.59 .\nn
\eea

From equation (\ref{hq1}), leading to $\lambda_1, \lambda_2, \lambda_4$ that must be 
 positive, satisfying the above bound conditions and 3.258 TeV$<m_{h_{++}}<$ 19.549 TeV . Thus the mass of the heavy Higgs must be some few TeVs, the electroweak phase
transition in this model is the first order phase transition. We hope that the
heavy particles will uncover many more new physics.
\section{Conclusion and outlooks}
\label{sec5} We have used the effective potential at finite
temperature to study the structure of the electroweak phase
transition in the RM331 model. This phase transition is separated
into two phases. The first transition period is $SU (3)\rightarrow
SU(2)$,  or the symmetry breaking in the energy scale $v_{\chi_0}$
 (in order to generate masses for the heavy
 particles and the exotic quarks). The second phase transition is
 $ SU (2)\rightarrow U(1)$ at $ v_{\rho_0} $, which generates masses for
 all usual fermions and the SM gauge bosons. The electroweak phase
  transition in this model (in the scale $v_\rho$) 
   may be the weakly first order electroweak phase
  transition with $m_{h_1}= 125$ GeV if the heavy bosons masses  are some
  few TeVs. So this is strong enough to study the baryon asymmetry.

If $ Z_2 $ exists, its mass is some few TeVs, so its contribution to the electroweak
phase transition is very large. Therefore, the electroweak phase transition in this
 model is completely turned  on. In other words,  the baryon asymmetry problem in this
 model is directly related to the mass of $Z_2$.

The self-interactions of Higgs in this model are more complicated
than the SM,  because heavy particles involve in  both two phase
transitions. Thus calculating the quantum corrections can reveal
many new physical phenomena  and opens up new relations further
between Cosmology and Particle Physics. In addition, from the
phase transitions, we can get some bounds on the Higgs
self-couplings.

Although we only work on the RM331 model, but this calculation can still apply to other
 3-3-1 models such as the recent  supersymmetric reduced minimal 3-3-1
 model \cite{srm331} and the 3-3-1-1 model \cite{3311}. So the 3-3-1 models can
  have specific advantages more than the SM in explaining the baryon asymmetry problem.

Our next works will calculate the sphaleron rate and CP violations in 3-3-1 models, 
in order to analysis  in details electroweak baryogenesis. 
In addition, by using the electroweak phase transition or baryogenesis problem,
 we can predict masses of the heavy particles beyond the SM such as 3-3-1 models.

\section*{Acknowledgment}
The authors  thank P. V. Dong and D. T. Huong for critical remarks and stimulating
discussions. This research is funded by Vietnam  National Foundation for
Science and Technology Development (NAFOSTED)  under grant number
103.01-2011.63.
\\[0.3cm]

\end{document}